# Information, uncertainty and holographic action

R.J. Dikken[1], K. Ng Wei Siang[1]

[1]Department of Materials Science and Engineering, DelftUniversity of Technology,
2628CD Delft, the Netherlands

**Abstract**

In this short note we show through simple derivation the explicit relation between information flow and the theories of the emergence of space-time and gravity, specifically for Newton's second law of motion. Next, in a rather straightforward derivation the Heisenberg uncertainty relation is uncovered from the universal bound on information flow. A relation between the universal bound on information flow and the change in bulk action is also shown to exist.

## I Introduction

One of the main goals of fundamental research is ultimately to develop simple and elegant formulations that describe nature. However, this proves to be difficult to achieve when attempts are made to reconcile the theory of general relativity and quantum theory. The theories developed, though elegant, are unfortunately not simple due to the high level of abstractness involved. Since the 1960s, various multi-dimensional theories with a deep mathematical basis were developed, for instance, the superstring theories [1]. These theories are different in their forms, and shown not to be general [1], but in fact are limit cases of a single theory, the M-theory [2].

String theory eventually led to the development of another abstract notion, the holographic principle. Proposed in 1993, this principle entails a dimensionality reduction theory that states that a 3+1 dimensional world can be actually encoded on 2+1 dimensional world [3]. Such dimensionality reduction puts restrictions to models that can explain quantum gravity. This theory was given a more explicit interpretation based on string theory in 1995 [4]. Three years later, the AdS/CFT correspondence [5] was proposed. The AdS/CFT correspondence gives the relation between the anti-de-Sitter spaces including quantum gravity expressed in M-theory, and the conformal field theories (quantum field theories) describing elementary particles.

With the holographic principle as basis, it was shown in 2011 that space-time and Newton's laws of gravity can be derived as emergent properties with entropic origin [6]. Thermodynamics seems to be at the heart of gravity, exemplified by the study of thermodynamics of black holes [7–10].

The connection between information and the associated complexity describing nature and computational complexity [11] implies also a direct relation with the universal measure of information flow. About three decades ago it has been theoretically proven that there exist a quantum limit to flow of information and entropy [12]. This is later experimentally verified with the measurement of the quantum of ther-

---

Corresponding author: R.J. Dikken,
robbertjandikken@hotmail.com

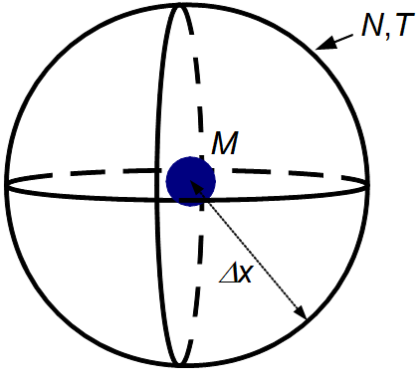

Figure 1: Schematic representation showing a mass enclosed by a holographic screen.

mal conductance [13]. Here, we aim to show, by simple derivation, the relation between the universal bound on information flow and emergent properties of space-time. Using the universal bound on information flow the uncertainty principle can be uncovered and a relation for bulk action [11] can be found.

## II  Newton's second law and information flow

From entropic principles, Newton's second law can be derived as an entropic force using a few simple assumptions [6]. Since the change of entropy is at the core of the emergence of the laws of inertia, and entropy is a measure of information space, it is hypothesized that the universal measure of the bound on information flow, or rather, the quantum of thermal conductance, is contained in the derivation in Ref. [6]. The maximum amount of information or energy that a channel can carry is given by

$$g_0 = \frac{\pi^2 k_B^2 T}{3h}, \qquad (1)$$

where $k_B$ is the Boltzmann constant, $T$ is the temperature and $h$ is the Planck constant. The quantum of thermal conductance is even more universal than for instance the conductance quantum of electronic transport, since $g_0$ is independent of the statistics of the carriers of information, while the conductance quantum of electronic transport applies only for Fermi-Dirac statistics. In the following we shall show that the entropic force is proportional to the quantum of thermal conductance $g_0$.

We start by defining a mass $M$ enclosed by a holographic screen with radius $\Delta x$. The associated number of bits (or channels) $N$ is related to the enclosing area, and thus related to $\Delta x$. The screen has a temperature $T$. A schematic representation is shown in Fig. 1.

Since Newton's second law can be derived as a emergent property from entropic principles [6], we start with Newton's second law and work towards the definition of the entropic force defined by the quantum of thermal conductance. The force $F$ is proportional to the acceleration $a$, with $M$ as proportionality constant:

$$F = Ma. \qquad (2)$$

Given that an enclosing screen is considered, the force of Eq. 2 is omni-directional. The energy $E$ related to the mass is given by the energy-mass equivalence

$$E = Mc^2. \qquad (3)$$

Through the equipartition theorem, this energy can be expressed in terms of screen properties

$$E = \frac{1}{2} N k_B T. \qquad (4)$$

Equating the two expressions for energy provides an explicit expression for the mass in terms of screen properties

$$M = \frac{1}{2} \frac{N k_B T}{c^2}. \qquad (5)$$

The relation between the temperature $T$ required to result in an acceleration $a$ is given by Unruh's law,

$$k_B T = \frac{1}{2\pi} \frac{\hbar a}{c}. \qquad (6)$$

Substituting Eq. 5 and the acceleration in Eq. 6 into Eq. 2, Newton's second law can be expressed as thermal energy of the screen,

$$F = \frac{1}{2}\frac{Nk_BT}{c^2}2\pi\frac{k_BTc}{\hbar}. \qquad (7)$$

Here, the reduced Planck constant, $\hbar = h/2\pi$, is replaced by the original Planck constant, since the quantum of thermal conductance is usually expressed in the original Planck constant. Simplifying the force expression and rearranging terms leads to

$$F = \frac{2}{c}N\frac{\pi^2 k_B^2 T}{h}T. \qquad (8)$$

Now we recognize the quantum of thermal conductance $g_0$ from Eq. 1 and we can express the force as

$$F = 6N\frac{g_0}{c}T. \qquad (9)$$

The force described by Newton's second law is related to information flow, given by a universal measure $g_0$. The number of bits $N$ represents the size of the area enclosing the mass $M$, and each bit can represent only a certain maximum amount of information. Interestingly it has been proposed that Hawking radiation and entropy flow from the horizon of a black hole can be considered as being produced in a one-dimensional Landauer transport process [14]. There, the quantum of thermal conductance is recovered, which relates the conductance of a single transport channel to fundamental constants, analogous to what is shown here.

## III  Heisenberg uncertainty relation and bound on information flow

Information can be thought of as analogous to (un)-certainty. One can express information flow by the change in energy expressed by entropy, $\Delta E = T\Delta S$, in time $\Delta t$:

$$J = T\frac{\Delta S}{\Delta t}. \qquad (10)$$

Given that the quantum of heat conductance puts a bound on information flow, this entails

$$J \leq \frac{\pi k_B^2 T}{6\hbar}T. \qquad (11)$$

Following Bekenstein [7], the change of information is given by $\Delta S = 2\pi k_B$. Substituting this in the above equation leads to

$$T\frac{2\pi k_B}{\Delta t} \leq \frac{\pi k_B^2 T}{6\hbar}T. \qquad (12)$$

With $\Delta E = 2\pi k_B T$, the Heisenberg uncertainty relation with a numerical pre-factor $1/\gamma$ is obtained:

$$\Delta E \Delta t \geq \frac{\hbar}{\gamma}. \qquad (13)$$

Although the physical meaning of the numerical prefactor $1/\gamma$ is unclear, it may be related to the geometry. Interestingly, following a straightforward reasoning, the Heisenberg uncertainty relation is uncovered from the bound on information flow. The bound on information flow may be more universal than the Heisenberg uncertainty relation, especially since the bound can be derived without considerations to the time/energy uncertainty principle [12].

## IV  Bulk action from information flow

Recently it is shown for a black hole how the change of bulk action is related to the associated bulk mass. It is conjectured that the complexity growth relates to the change in action, and that the complexity $C$ relates to action $A$ by $C = A/\pi\hbar$ [11]. Using the found relation, in Section II, between the entropic force and information flow, we derive here an expression for the change in action similar to ref. [11]. The purpose is to investigate the relation between the bound of complexity growth and the universal bound on information flow, the quantum of thermal conductance. Following the same conjecture as in Ref. [11], that complexity is proportional to action, we can focus on a bound for the change of action as a measure for the complexity growth.

The entropic force is defined by the change in entropy $\Delta S$

$$F = T\frac{\Delta S}{\Delta x}. \quad (14)$$

Since Eq. 2 is derived from Eq. 14, we equate Eq. 14 with Eq. 7

$$T\frac{\Delta S}{\Delta x} = N\frac{\pi k_B^2 T}{\hbar c}T. \quad (15)$$

Next, we use the equipartition rule to express the thermal energy $k_B T$ in the mass $M$

$$k_B T = \frac{2Mc^2}{N}. \quad (16)$$

If we insert this thermal energy in Eq. 15 we find

$$\frac{\Delta S}{\Delta x} = \frac{2\pi Mck_B}{\hbar}. \quad (17)$$

The proportionality between entropy and action $S = k_B A/\hbar$ [15] entails that a variation in entropy relates to a variation in action

$$\Delta S = \frac{k_B \Delta A}{\hbar}. \quad (18)$$

Substituting Eq. 18 in Eq. 17 gives

$$\frac{k_B}{\hbar}\frac{\Delta A}{\Delta x} = \frac{2\pi Mck_B}{\hbar}. \quad (19)$$

Since the holographic screen (or black hole horizon) is defined by light-like geodesics, the metric is $\Delta s^2 = 0$ and hence $\Delta x^2 = c^2 \Delta t^2$. This substitution, assuming infinitesimal variation, leads to the expression for the change in action similar to the result found in [11]:

$$\frac{dA}{dt} = 2\pi Mc^2. \quad (20)$$

The derivation in [11] to reach the result of Eq. 20 is far from trivial and involves a complicated cancellation between the Einstein-Hilbert volume term and the York-Gibbons-Hawkings surface term in the action integral. The derivation followed here is simpler and entails the same result. However, it remains a question what meaning such derivation has in terms of physical reality.

## V Conclusions

In this work we have derived, in a simpler way, the relation between the universal bound on information flow, or rather, the quantum of thermal conductance, and the entropic force that represents Newton's second law of motion. It is shown in a rather straightforward fashion that the Heisenberg uncertainty relation can be derived from the fundamental bound on information flow. Lastly, the information flow is directly related to the change of bulk action, and hence, by the conjecture in Ref. [11], also to the growth of complexity. Interestingly, simple derivations show relations that seem to have deep fundamental meaning. However, whether these relations entail physical reality, or that the relations are of apparent nature, remains a subject of discussion.